\documentclass[11pt]{article}

\usepackage{natbib}
\usepackage{caption,geometry}
\usepackage{epigraph}
\usepackage{anysize} 
\marginsize{2.5cm}{2.5cm}{3cm}{3cm}
\usepackage[final]{pdfpages}
\usepackage[utf8]{inputenc}
\usepackage[T1]{fontenc}
\usepackage{amsmath}

\usepackage{amsmath}

\usepackage{style/lettrine}
\usepackage{style/overrulehere}
\usepackage[Lenny]{style/fncychap}
\usepackage{style/floatfig}%
\usepackage{style/supertabular}%
\usepackage{style/multirow}
\usepackage{style/subfigure}
\usepackage{style/soul}
\usepackage{style/emptypage}

\usepackage[resetlabels]{multibib}

\def\calwf3{{\tt{calwf3}}}

\let\deg=\circ

\def\lesssim{\mathrel{\hbox{\rlap{\hbox{\lower4pt\hbox{$\sim$}}}\hbox{$<$}}}}
\def\gtrsim{\mathrel{\hbox{\rlap{\hbox{\lower4pt\hbox{$\sim$}}}\hbox{$>$}}}}

\def\d{{\rm d}}

\def\deg{\ifmmode^\circ\else$^\circ$\fi}
\def\alphaTF{\ifmmode{\alpha_{\mathrm{\,{\small TF}}}}\else{$\alpha_{\mathrm{\,{\small TF}}}$}\fi}

\def\Msun{\ifmmode{\mathrm M_\odot}\else{M$_\odot$}\fi}

\usepackage{authblk}
\begin{document}

\title{\textbf{A historical perspective on the concept of galaxy size}\thanks{Based on my Ph.D. thesis at IAC/ULL.}}



\author[1, 2]{Nushkia Chamba\thanks{Now moved to Stockholm University/Oskar Klein Centre. Email: \texttt{nushkia.chamba@astro.su.se}}}
\affil[1]{Instituto de Astrof\'isica de Canarias, c/ V\'ia L\'actea s/n, E-38205, La Laguna, Tenerife, Spain}
\affil[2]{Departamento de Astrof\'isica, Universidad de La Laguna, E-38205 La Laguna, Tenerife, Spain}

\date{}

\maketitle

\begin{abstract}
    A brief narrative on how the effective radius and isophotal diameters were accepted as galaxy size measures is presented. Evidence suggests that these parameters were defined only based on observational premises, independent of any astrophysical theories. An alternative, new physically motivated size definition based on the expected gas density threshold required for star formation in galaxies is discussed. The intrinsic scatter of the size--stellar mass relation using the new size measure is 0.06\,dex, three times smaller than that of the relation with the effective radius as size. The new physically motivated size measure can be adopted in upcoming deep, wide imaging surveys. 
\end{abstract}

\small{\textit{Keywords}: Galaxy evolution (594); Galaxy formation (595); Galaxy properties (615); History of astronomy (1868)}


\section{How the concept of galaxy size evolved}

\label{ch1:sec:size_history}

Prior to the popularisation of the effective radius \citep[$R_{\rm e}$;][]{1948AnAp...11..247D} as a galaxy size measurement, it was common to refer to the extensions of galaxies as `diameter' or `dimension' in the literature. The apparent diameter of galaxies was initially discussed by \citet{1926hubble} and in later works `diameter' was defined using the location of surface brightness isophotes. \citet{1936MNRAS..96..588R} was the first to suggest such a definition as a `conventional' measure for the diameter of elliptical `nebulae'. He showed that the measurements of the angular diameters for six elliptical galaxies made by \citet{1926hubble} and \citet{1934shapley} were dependent on observing conditions (instrumental and atmospheric) under which the photographic plates were obtained. Therefore, the measurements from these earlier works were difficult to reproduce, owing to the absence of a clear boundary for the galaxies. For these reasons, Redman proposed the use of a fixed surface brightness isophote at 25\,mag/arcsec$^2$ as a definition of the diameter for elliptical galaxies (relevant text highlighted here in bold in the following excerpt):

\begin{quote}
    `\textit{Enough has been said to make it clear how extremely
difficult it is to discuss questions relating to the diameters or mean surface
brightnesses of these objects...\textbf{There is in fact no
boundary to any of these nebulae; they all shade off imperceptibly into empty space, so that we can only speak of a ``diameter'' by defining some
conventional boundary}, much as we have a conventional definition for the
``half-width'' of a spectrum line, another entity which is not definitely
bounded. \[...\] The best alternative is to \textbf{define the boundary as the locus of points at which the surface brightness has a certain value in stellar magnitudes per
square second of arc...the mean limit is 25$^{\rm m}$.3/square second of arc, which is practically identical
with} the limit 25$^{\rm m}$.2/sec.$^2$
found by \citet{1932hubble} for \textbf{the faintest surface brightness detectable on certain photographs} (70--90 minutes exposure) taken with
the 60-inch reflector at Mount Wilson. Hubble concludes that a reasonable
limit for prolonged exposures under the best conditions is probably between
26$^{\rm m}$.5/sec.$^2$ and 27$^{\rm m}$.0/sec.$^2$ -- This applies to Mount Wilson ; the...limit [of photographs from the Solar Physics Observatory in Cambridge] would not be quite so faint. \textbf{It seems that a conventional boundary at 25$^{\rm m}$/square second of arc would have the following advantages: (a) It would allow a definite measure of the ``diameter'' of the nebula ; (b) it can
be attained on photographs made in an indifferent climate ; (3) it would
include most of the light of the nebula.}}'\\
\hspace*{0pt}\hfill ---\citet{1936MNRAS..96..588R}
\end{quote}

In other words, the specific 25\,mag/arcsec$^2$ isophote or `boundary' chosen by Redman stemmed from two facts: 1) it was approximately the mean limiting depth of photographic plates from the Solar Physics Observatory and 2) it enclosed the bulk of the light distribution of the galaxies he studied. It should be noted, however, that \citet{1936MNRAS..96..588R} used photograhs of only five elliptical galaxies (out of the six in the sample considered) to propose this definition of `diameter of the nebula'. Hereafter, the isophote at 25\,mag/arcsec$^2$ in $B$-band (parameterised by the major-axis of an ellipse) will be denoted as $D_{25}$, following the notation in the well-known Second and Third Reference Catalogue of Galaxies \citep[RC2 and RC3;][]{1976RC2...C......0D, 1991dev}.\footnote{$D_{25}$ is frequently (and incorrectly) referred to as $R_{25}$ in the literature. In RC2 (and RC3), $R_{25}$ is the \textit{ratio} between the major and minor axes at the location of the 25\,mag/arcsec$^2$ isophote in $B$-band.}

A second isophotal measure for diameter emerged two decades later when \citet{1958MeLuS.136....1H} heroically extracted microphotometer tracings of three hundred nebulae imaged using 103a photographic plates\footnote{The 103a photographic plate series were less sensitive to low signal-to-noise sources compared to the IIIa plates in the late 1960s (which were used by  \citet{arp1969} to detect the first stellar halo of a galaxy).}, observed mostly from Mount Wilson between 1947--55. Holmberg sought to measure accurate total magnitudes, colour indices and apparent dimensions of galaxies distributed in the Northern Sky. He defines `diameter' as the radial location in a galaxy where the photographic density with respect to the background in the plate is $0.5$\%, noting that:

\begin{quote}
    `\textit{[A] relative plate density of $0.5\%$ corresponds, on an average, to surface magnitudes of 26$^{\rm m}$.5 (photogr. reg.) and 26$^{\rm m}$.0 (photogv. reg.), per square second. Since a density of $0.5\%$ is close to the practical measuring limit, the two magnitudes mentioned represents the limiting surface magnitudes of the plates.}'\\
\hspace*{0pt}\hfill ---\citet{1958MeLuS.136....1H}
\end{quote}

\noindent Therefore, while the limiting surface brightness magnitudes in the collection of photographic plates used by \citet{1958MeLuS.136....1H} was 26.5\,mag/arcsec$^2$, the \textit{definition} of `diameter' was based on a relative density of $0.5\%$ in the sample of galaxies with respect to the background in the plates. Indeed, such a statistical definition of diameter depends on the sample of galaxies and the limiting depth of the images, which would then correspond to different surface brightness isophotes. For operational purposes, however, Holmberg's radius ($R_{\rm H}$) has been measured using the location of the isophote at 26.5\,mag/arcsec$^2$ in the $B$-band \citep[which is close to the photographic band used by][]{1958MeLuS.136....1H}.\footnote{See \citet[][Appendix A]{2007vanderkruit} for a discussion on the precise definition of Holmberg's radius, as well as a demonstration of the remarkable reproducibility of the measurements by \citet{1958MeLuS.136....1H} despite the many difficulties he faced in tracing plates. It is also noteworthy that \citet{1958MeLuS.136....1H} measured the properties of several low surface brightness galaxies (with a mean surface brightness $< 26\,$mag/arcsec$^2$).} \par  



From the above discussion, it is clear that both $D_{25}$ and $R_{\rm H}$ were operatively selected by \citet{1936MNRAS..96..588R} and \citet{1958MeLuS.136....1H}, respectively, to measure the maximum apparent boundaries of galaxies. While it seems as though de Vaucouleurs advocated the use of $R_{\rm e}$ as a measure of galaxy dimension throughout his career\footnote{All of his publications referenced in this article, but I also refer the reader to the International Astronomical Union Symposium (IAUS) proceedings \citet{1974IAUSdev} and \citet{1987IAUSdev}.}, he frequently included the isophotal measures in galaxy catalogues ($D_{25}$ in RC2 and RC3) and in studies where he consistently adopted the terms `diameter' and `dimension' interchangably to refer to these parameters \citep[including $R_{\rm e}$, see also e.g.][]{1959devucouleursb}. However, the use of `diameter' and `dimension' interchangably in the literature to describe the concept of galaxy size was evident as early as the 1930s--40s, i.e. before \citet{1948AnAp...11..247D} introduced $R_{\rm e}$. This was a period when long exposures and homogeneous plates were increasingly used to quantitatively compare the dimensions of spheroidal and spiral galaxies\footnote{This was an interesting issue in that era of extragalactic observations because reflector or refractor plates showed that spiral galaxies were significantly larger (by a factor five) than ellipticals, and astronomers were interested in looking for `growth' in dimensions along the Hubble sequence (\citet{1942shapley} but see also \citet{1959devucouleursb} and references therein).}: 

\begin{quote}
    `\textit{Since galaxies, like planetary atmospheres, probably fade out indefinitely, the extreme \textbf{diameters} are not very important in themselves (if the peripheral masses are negligible); but the \textbf{dimensions} out to a given light or mass density become very significant if evolutionary trends are under
consideration and comparative \textbf{sizes} and gradients must be discussed.
\[...\] A by-product of Miss [Dr] Patterson's photometry [using homogenous, long exposure plates from Havard Observatories] has been the measurement of \textbf{boundaries}. From this material, the intercomparison of dimensions of spheroidal and spiral galaxies now becomes possible...and is, in fact, about the first to be precisely suited [to address this problem].}'\\
\hspace*{0pt}\hfill ---\citet{1942shapley}
\end{quote}

\noindent From the above excerpt from \citet{1942shapley}, it is quite clear that `diameters', `dimensions' and `boundaries' (highlighted here in bold for emphasis) all refer to the sizes of galaxies. The fact that many different terms are used to describe the same concept suggests that a clear framework to define exactly what `size' meant for a galaxy had not yet been realised. This is probably because the outer regions of galaxies were poorly understood in that era (this was recognised) and that observations were highly biased towards the brightest regions of galaxies \citep{1976disney}. Regardless of these issues, even in these early discussions, astronomers were concerned about the right method of comparing the relative diameters or sizes of galaxies to study their growth and evolution. Shapley was well aware that only long-exposure, homogenous datasets (i.e. deep imaging) can be used to compare the sizes of different galaxy populations. Furthermore, he recognised that if sizes of different populations are to be compared in the context of evolutionary scenarios, then the choice of the dimension parameter is extremely relevant.\par
Given the situation regarding the nomenclature and description of galaxy size in the 1930--40s (and even after de Vaucouleurs introduced $R_{\rm e}$ in the 1950s), it begs the question of when did astronomers start referring to parameters such as $R_{\rm e}$, $D_{25}$ and $R_{\rm H}$ as size? At least in the case of $R_{\rm e}$, this seems to have happened slowly and gradually, starting around the 1960s--70s, after the works by \citet{1959devucouleursb}, \citet{1963fish} and \citet{1968sersicb}. The details of how this may have happened will now be described. \par 

After $R_{\rm e}$ was introduced by \citet{1948AnAp...11..247D} in the form of the $R^{1/4}$ model for elliptical galaxies given by:

\begin{equation}
    \mu(R) = \mu(0)e^{-b(R/R_{\rm e})^{1/4}},
\label{ch1:eq:r4_law}
\end{equation}

\noindent many astronomers agreed that it was a universal and physical law for these galaxies. A significant number of investigations were thus dedicated to understand what the $R^{1/4}$ law implied for the formation of elliptical galaxies.  \citet{1963fish} was the first to demonstrate a correlation between de Vaucouleurs' $R_{\rm e}$ and absolute luminosity $L$ for elliptical galaxies in the Virgo cluster, $L \sim R_{\rm e}^{3/2}$. He called this the `Luminosity Concentration Law': probably because $R_{\rm e}$ is a measurement of light concentration\footnote{At a fixed stellar mass, galaxies with a smaller value of $R_{\rm e}$ are more concentrated than galaxies with larger $R_{\rm e}$. However, more quantitative measures of galaxy light concentration are $R_{\rm e}/M$ or $R_{\rm e}/L$.} in galaxies and a `law' because he attempted to interpret the correlation in terms of a model for galaxy formation. Fish expanded on these results a year later, by studying the correlation between the mass and potential energy of elliptical galaxies i.e. $R_{\rm e} \propto M^{\alpha}$:

\begin{quote}
    `\textit{The significance of a relationship between mass and potential energy in elliptical
galaxies lies in two facts. First, the relationship describes the \textbf{degree of concentration}
of the elliptical galaxies at the time of star formation. Second, it reveals the dependence
of the radiation upon total mass during the condensation of the protogalaxy out of the
pregalactic medium. The first fact permits drawing conclusions about conditions for
star formation. The second allows selection between the physical events that might have
occurred during the contraction of the protogalaxy. \[...\] 
(A relationship between
mass and angular momentum in elliptical galaxies would provide further information
on the condensation process, but no significant correlation is found to exist
between these two parameters.)}'\\
\hspace*{0pt}\hfill ---\citet{1964fish}
\end{quote}

\noindent In other words, while it was possible to make a physical interpretation of the $R_{\rm e}$--mass or $R_{\rm e}$--luminosity correlations using galaxy formation ideas in that period, considering $R_{\rm e}$ as a measure of the `degree of concentration' (highlighted here in bold) for elliptical galaxies,  a relationship with angular momentum of the galaxy was still far from being understood. Apart from the above physical interpretations he lays out in the article, \citet{1964fish} also notes that de Vaucouleurs' $R_{\rm e}$ is much more reliable to obtain compared to isophotal radii that demanded accurate photometry and knowledge of absolute values of surface brightness. Rewriting de Vaucouleurs law as:

\begin{equation}
    \mu(R) = \mu_e e^{-7.76[(R/R_{\rm e})^{1/4} -1]}
\end{equation}

\noindent where $\mu_e = \mu(R_{\rm e})$, one can readily show that 

\begin{equation}
    \log{\left(\frac{\mu(R)}{\mu(0)} \right)} = -3.37 \left(\frac{R}{R_{\rm e}} \right)^{1/4}.
\end{equation}

Therefore, by simply plotting $\mu(R)$ vs. $R^{1/4}$, it is possible to determine $R_{\rm e}$ from the slope of the relation, provided an accurate linear regression:

\begin{quote}
    `\textit{This fact reveals an advantage of the de Vaucouleurs definition of radius over an
isophotal definition, because an isophotal radius can be no more reliable than the accuracy
with which the zero point of the photometry is determined. Good zero points require
fairly elaborate procedures in surface photometry, particularly if the technique is completely
photographic. It should also be mentioned that the simple form for the potential
energy given by equation (8) [$\Omega \propto GM/R_{\rm e}$] is not valid for a radius defined simply as the distance from the nucleus to a given isophote.}'\\
\hspace*{0pt}\hfill ---\citet{1964fish}
\end{quote}

\noindent What this means is that Fish preferred $R_{\rm e}$ because it was more robust compared to measurements of isophotal radii using the photographic plates of that time and also because $R_{\rm e}$ generated very simple and elegant equations for physical quantities, like potential energy $\Omega$. Therefore, given the limitations in imaging in the 1960s as well as the difficulty in simulating galaxy formation theories, it is completely justifiable that a lot of astronomers used $R_{\rm e}$ to analyse the formation of elliptical galaxies in these times. However, even in Fish's paper, $R_{\rm e}$ was never perceived (or explicitly associated) as a diameter or size definition for galaxies, but as a measurement of concentration and `radius' because it has units of length. \par

Four years later, \citet{1968adga.book.....S} made similar efforts in the understanding of the formation of elliptical galaxies by generalising de Vaucouleurs $R^{1/4}$ law using an index `$n$':

\begin{equation}
\label{ch1:eq:sersic}
    \mu(R) = \mu(0) e^{-b_n(R/R_{\rm e})^{1/n}}.
\end{equation}

\noindent This is called the S\'ersic law. Setting $n=4$ and $n=1$ gives back Eq. \ref{ch1:eq:r4_law} (the $R^{1/4}$ law for ellipticals) and an exponential for spirals, respectively. S\'ersic was well aware that not all elliptical galaxies followed the $R^{1/4}$ law.\footnote{Some researchers would even tweak the background level in imaging just to force galaxies under study to the $R^{1/4}$ model \citep[pointed out by][]{1994donofrio}.} He subsequently studied the `mass--radius diagram', i.e. the mass--$R_{\rm e}$ relationship, for a sample of early and late-type galaxies that same year \citep{1968sersicb}. He included late-type spiral galaxies in his sample partly because \citet{1964fish} associated the elliptical galaxies he characterised with neighbouring spirals to ascertain whether spirals were `youthful' or not. \par
\begin{figure}[h]
    \centering
    \includegraphics[width=0.6\textwidth]{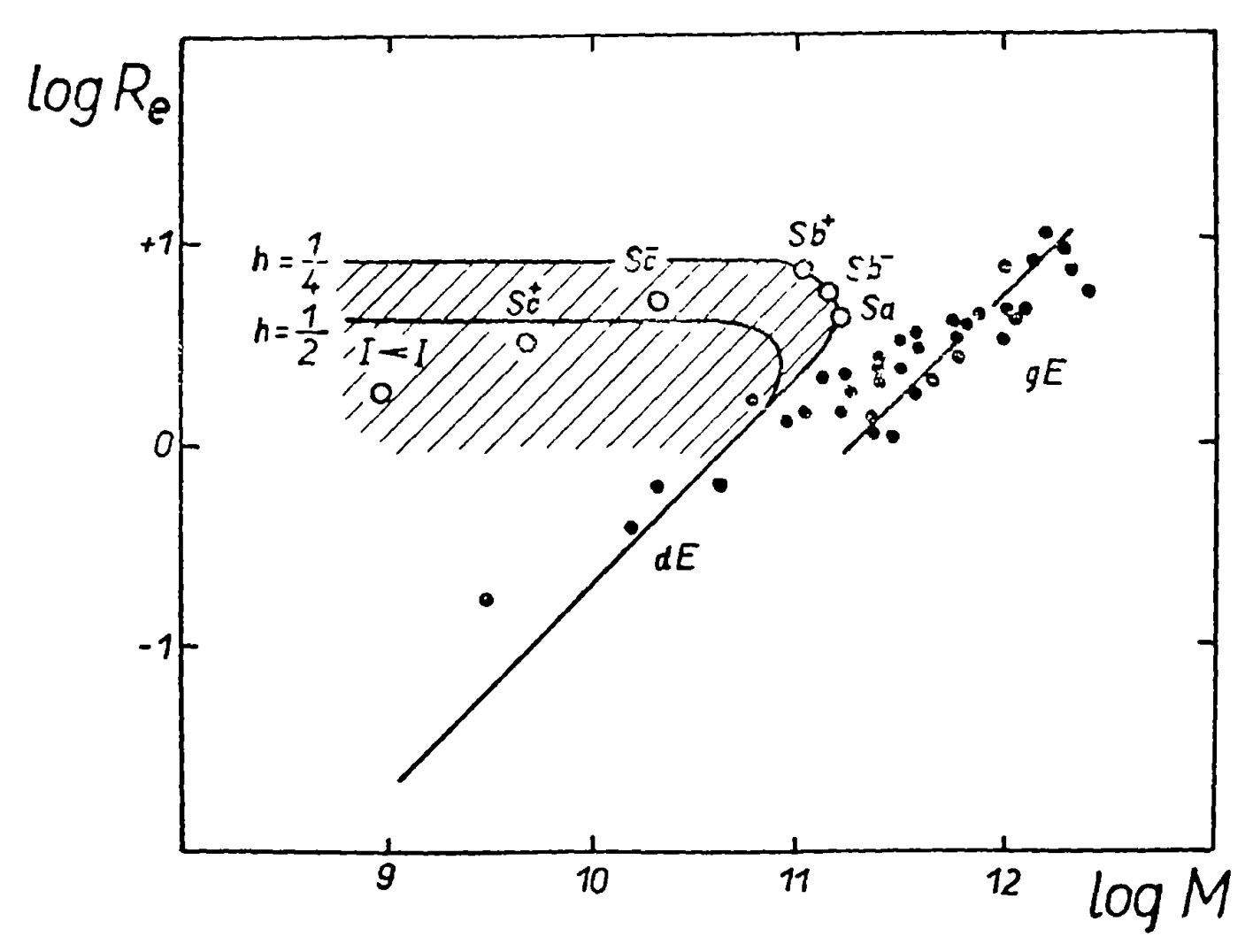}
    \caption{The first mass--radius diagram for galaxies. The shaded region suggests the qualitative dispersion of the individual data points. The spiral (open circles), elliptical (black circles) and dwarf elliptical (dE) galaxies are clearly marked, already showing hints of the different sequences that these galaxy types populate in this parameter space. \textit{Credit: \citet{1968sersicb}}.}
    \label{ch1:fig:mass-radius-sersic}
\end{figure}

 
S\'ersic's diagram is shown in Fig. \ref{ch1:fig:mass-radius-sersic}. Even though he used a relatively small sample, recognising selection effects and an over-representation of interacting galaxies (which is probably why he did not attempt to quantify the slope or dispersion of the relation), S\'ersic's diagram already begins to reflect two distinct sequences: an almost flat region in the $R_{\rm e}$--mass plane for late-type galaxies and a separate sequence for the ellipticals. Like those before him, S\'ersic was interested in trying to find out whether an evolutionary sequence occurred along these Hubble sequences. For this reason, he interpreted the region around $10^{11}\,M_{\odot}$ as a `transition region' between the late and early-types because it was dominated by the interacting galaxies in his sample.

It is of importance to point out that the $R_{\rm e}$ for the spiral galaxies studied by \citet{1968sersicb} was computed using the calibrated, linear relation between $R_{\rm e}$ and $D_{25}$ from \citet{1959devucouleursb}, and not by fitting an exponential model to the profiles of spirals. In fact, the attempt by \citet{1959devucouleursb} can be considered the first to compare different measures of the parameters we call size today, i.e. $R_{\rm e}$ and $D_{25}$. It was also during this time that de Vaucouleurs promoted the `effective radius' to an `effective dimension', essentially placing it and the isophotal diameters (or `brightness dimensions' as he called them) on the same footing. This could be an early source for the confusion in the literature between the original notion of dimension or diameter \citep[as discussed in e.g.][]{1926hubble, 1936MNRAS..96..588R, 1942shapley} and what the effective radius was tracing, i.e. light concentration \citep[e.g. in][]{1963fish}. Subsequent comparisons were conducted by \citet{1961vorontsov} and  \citet{1970genkin}. But although these comparisons have been completely ignored to date, it turns out that they provide further insight in to the development towards referring to parameters like $R_{\rm e}$, $D_{25}$ or $R_{\rm H}$ as `size'. In light of this, I highlight some of the discussion in the work by \citet{1961vorontsov} and  \citet{1970genkin} below. \par 

\section{A tale of two citations}

Although astronomers had devoted a lot of attention towards the use of the de Vaucouleurs law in the early 1960s, there is little evidence to suggest that most of these researchers had considered the negative implications (if any) on the use of $R_{\rm e}$ for the understanding of elliptical galaxies, as well as of galaxies of other types. Additionally, although the dimensions or diameters of galaxies had indeed been discussed in the context of $R_{\rm e}$, $D_{25}$ and $R_{\rm H}$, the concept of `size' itself was not yet explicitly defined in these early works --- it seemed as though it was continuously assumed that they all could be used to describe galaxy size and for any galaxy type. On this aspect, \citet{1961vorontsov} calls for a more explicit definition of the dimension of galaxies and offers a critical perspective on the use of diameters to compare elliptical and spiral galaxies as performed in the work of de Vaucouleurs:


\begin{quote}
     `\textit{Different methods of determination of the diameters of spiral galaxies give results which are principally not comparable. It is proposed that diameters of distinctly visible spiral patterns be compared. Their diameters depend only very slightly on exposures, etc. The definition of diameters of elliptical galaxies to be used for comparison is less obvious. \textbf{At the present stage, the comparison of diameters} of spiral and elliptical \textbf{galaxies has no sense until the physical and practical determination of the boundaries of galaxies has been properly adjusted}.}'\\\\
\hspace*{0pt}\hfill ---\citet{1961vorontsov}
\end{quote}

\noindent In other words, the determination of diameters for ellipticals and spirals should be performed separately because of their different internal structure, and for this reason comparisons are difficult if not impossible, unless a more physically meaningful and practical approach is adopted (in bold here for emphasis). However, in the early 1960s, finding such a definition of diameter or size was not possible as the outskirts of galaxies were unknown \citep[][recognised this]{1961vorontsov}, at least not until the first detection of stellar haloes \citep{devaucouleurs1969, arp1969}. \par 
The second of these works, \citet{1970genkin}\footnote{As of 15 Oct. 2020, this is the only other article which cites \citet{1961vorontsov}.}, seems to be the first ever paper to have `size' in its title, i.e. `\textit{Comparisons of Galaxy Sizes}', referring to the `effective diameter' and isophotal diameters. Like \citet{1959devucouleursb} \citep[also in][]{observatoireannales}, these authors were interested in establishing a statistical relationship between the different size definitions for galaxies. However, they also correctly point out that $R_{\rm e}$ is not very well suited for spiral galaxies, while it is more appreciable for elliptical galaxies (highlighted here in bold):

\begin{quote}
    `\textit{In general a determination of effective dimensions does not appear to be of any value at all for spiral galaxies. On the one hand, the \textbf{effective sizes are not related to any structural features} of spiral systems, while on the other, [...] \textbf{[f]or elliptical galaxies having no structural features, the effective size is a very convenient parameter}, although it remains difficult to determine.}'\\\\
    \hspace*{0pt}\hfill ---\citet{1970genkin}
\end{quote}

\noindent In other words, echoing \citet{1961vorontsov}, $R_{\rm e}$ is not related to any physical characteristics of galaxies, it is not physically\footnote{By physical, I mean a property innately linked to how galaxies grow and evolve in size.} motivated. Additionally, as $R_{\rm e}$ was primarily introduced via a model to parameterize the light distribution for elliptical galaxies specifically \citep{1948AnAp...11..247D}, $R_{\rm e}$ is not valuable for measuring the sizes of spiral galaxies. \par 

Although both the above works recognised the limitations of using $R_{\rm e}$ very early on, $R_{\rm e}$ continues to be the top choice among astronomers for characterising galaxy size. One possible reason for the popularity of $R_{\rm e}$ could be its use in several classical works on massive elliptical galaxies from the 1970s--80s \citep[e.g.][]{1977kormendy, 1987DS}. In these works, $R_{\rm e}$ was used in various scaling relations that were later shown to produce constraints on galaxy formation models for elliptical galaxies. A closer look at \citet{1977kormendy}, however, is interesting because the author showed that the \citet{1930hubble}, \citet{1948AnAp...11..247D} and \citet{1962king} models for spheroidal galaxies  `\textit{all measure essentially the same physics. Therefore, we are free, in the following discussion, to
use the most convenient fitting function, namely, the
de Vaucouleurs (1948, 1953) law}'. In other words, while the physics underlying the galaxy is independent of $R_{\rm e}$ specifically, Kormendy argued that de Vaucouleurs law is a convenient fitting function for spheroidal light profiles.\footnote{In more recent work, however, where \cite{1968adga.book.....S} law is used to model galaxy light profiles, $R_{\rm e}$ is computed numerically.} \par \bigskip 

In the above historical discussion, we have seen that while $R_{\rm e}$ can be used to understand the formation of galaxies via for example, the Virial Theorem \citep{1964fish} and the `mass--radius' diagram \citep{1968sersicb}, the popular use of $R_{\rm e}$ also evolved the concept of galaxy size adopted in the literature. In the 1930s--40s, galaxy size was predominantly described in terms of galaxy boundaries \citep[e.g.][]{1936MNRAS..96..588R, 1942shapley} and later in the 1950s, $R_{\rm e}$ was promoted as a parameter to measure galaxy size \citep[e.g.][]{1959devucouleursb}, even though $R_{\rm e}$ was not specially introduced by \cite{1948AnAp...11..247D} for this purpose. In the following sections of this article, I briefly discuss the implications of $R_{\rm e}$'s popularity on our understanding of the sizes of galaxies in modern astronomy and motivate the need for a more physically representative definition of galaxy size which is related to the concept of galaxy boundaries.






\section{Implications of $R_{\rm e}$'s popularity}
\label{ch1:sec:thesis_overview}

In this work, I have discussed the historical origins of the popularity of $R_{\rm e}$ and other classical parameters such as $D_{25}$ as a galaxy size measure. In particular, I showed that parameters such as $R_{\rm e}$ and the isophotal radii have always been assumed to represent galaxy size, but never explicitly demonstrated in these terms within a clear physical framework. $R_{\rm e}$ was specifically chosen to parameterise the $R^{1/4}$ model for elliptical galaxies. \citet{1948AnAp...11..247D} fixed the scaling radius $R_{\rm s} \equiv R_{\rm e}$ in Eq. \ref{ch1:eq:r4_law} probably because galaxy outskirts were not well observed using photographic plates at that time. However, in principle, \textit{any} $R_{\rm s}$ (i.e. radii enclosing different fractions of total galaxy light) could have been used for this purpose and therefore this makes $R_{\rm e}$ an arbitrary parameter that has consequences on the interpretation of scaling relations \citep[for a discussion on this issue see][]{2019PASA...36...35G}. \par 

As mentioned previously, $R_{\rm e}$ was preferred among many astronomers in the past primarily because the outskirts of galaxies were not well understood or observed with low enough uncertainty \citep[at least not until the low signal-to-noise IIIa-J plates were improved in the 1970s and stacked to produce deep images; see e.g.][]{1999malin}. To some extent, this is still especially true for very high redshift galaxies. This is because observational conditions or limitations prevent the outer regions of high-redshift galaxies to be measured accurately due to, for example, cosmological dimming by a factor $\propto (1 + z)^4$. This has potentially influenced how astronomers view the use of $R_{\rm e}$ as a size measure for galaxies because $R_{\rm e}$ is practically more advantageous under these circumstances. 


But at the same time, in the case of nearby galaxies, even when their outskirts were being observed with much greater accuracy, starting with massive surveys like SDSS \citep[][]{2000york}, the association of $R_{\rm e}$ to galaxy size had already been so deeply established in the astronomical literature that there seemed to be no room for criticism. In fact, the first statistical exploration of the mass--size relation in the nearby Universe by \citet[][]{2003shen} made use of SDSS, which catalogued $R_{\rm e}$ for a million galaxies in the late 1990s--early 2000s. Therefore, subsequent studies in the last 15--20 years probing the size evolution of galaxies at higher redshift were practically forced to select $R_{\rm e}$ to represent galaxy size for consistency, because the work by \citet{2003shen} was a reference for objects at low redshift. In other words, studies using $R_{\rm e}$ to represent the sizes of nearby galaxies (since the works by e.g. \citet{1959devucouleursb}, \citet{1963fish}, \citet{1968sersicb} and later \citet{2003shen}) have influenced studies at high redshift afterwards, eventually leading to a convention that $R_{\rm e}$ is a measure of galaxy size. Consequently, what this means is that the current consensus on the size evolution of galaxies is in reality a consensus on their $R_{\rm e}$ evolution. \par 

Investigations attempting to study the mass--size relation have also used radii enclosing other fractions of the total galaxy light, e.g. $R_{90}$ that encloses 90\% of the galaxy light \citep[][]{2011nair} or $R_{80}$ that encloses 80\% \citep[][]{2019miller}.
But only a few works in the literature have explored the use of other indicators for galaxy size that are completely different in nature to $R_{\rm e}$ or other light concentration measures. These include the break radii for disc galaxies studied at low and moderately high redshift  \citep[$z \sim 1$;][]{2005ApJ...630L..17T, 2008ApJ...684.1026A} and various isophotal measures \citep{2011ApJ...726...77S, 2012MNRAS.425.2741H, 2015ApJS..219....3M}. Interestingly, unlike $D_{25}$ which was introduced using elliptical nebulae, more recent studies have used isophotal measures to represent galaxy size for disc galaxies, e.g. $R_{23.5, i}$ which is the radius where $\mu = 23.5$\,mag/arcsec$^2$ in the optical SDSS $i$-band \citep{2012MNRAS.425.2741H} as well as the isophote at 25.5\,mag/arcsec$^2$ in the 3.6$\mu$ near-infrared imaging from the Spitzer Survery of Stellar Structure in Galaxies (S4G)
survey \citep[]{2010PASP..122.1397S, 2015ApJS..219....3M}. All the above light concentration and isophotal size measures are somewhat arbitrary in the sense that they do not necessarily hold any physical information on the formation of galaxies. However, thus far, it is $R_{\rm e}$ (or variants such was $r_{1/2}$) that has been used in important theoretical studies on the connection between size, mass and angular momentum of galaxies \citep{1998MNRAS.295..319M, 2008somerville, 2013ApJ...764L..31K, 2019MNRAS.tmp.1977J} and in the calibration of current hydrodynamical simulations, e.g. in EAGLE \citep{2015crain} and IllustrisTNG \citep{2018pillepicha}, among others. \par

\section{Towards a physically motivated definition of galaxy size}

In light of the above situation, in \citet*{2020tck} we introduced a new, physically motivated definition for the size of a galaxy using deep imaging. The work is not only timely, given knowledge on the outskirts of galaxies and their formation, but necessary as until now, an explicit, more physical and practical definition of the luminous size for galaxies has barely been explored. Additionally,  none of the current size measures in the literature have been critically analysed to reaffirm their suitability to compare the sizes of galaxies of different morphology. This is important for most, if not all, of the classical size measures (e.g. $R_{\rm e}$ and $D_{25}$) defined using massive elliptical galaxies as well as those parameters which were later explored specifically for spiral galaxies \citep[e.g.][]{2005ApJ...630L..17T, 2012MNRAS.425.2741H}. In fact, to some extent, we are still very much in a similar situation as that pointed out in \citet{1961vorontsov} all those decades ago: how can the diameters of different galaxy types be fairly compared within a physical framework? This question is addressed in \citet*{chamba2020}, who point out that the use of $R_{\rm e}$ to compare the sizes of galaxies can be misleading. An example is the statement that ultra-diffuse galaxies are `Milky Way-sized' \citep{2015giantgalaxies}.  \par 

The new physically motivated size definition is based on the expected gas density threshold required for star formation in galaxies. In \citet{2020tck}, we selected a stellar mass density threshold as a \textit{proxy} for such a definition. We chose the radius at the 1\,$M_{\odot}/$pc$^2$ isomass contour (which we called $R_1$) because it indicates the density at the truncation in Milky Way-like galaxies \citep[see][]{2019MNRAS.483..664M}. We measured $R_1$ for $\sim 1000$ galaxies, from dwarfs to ellipticals, with stellar masses between $10^{7}\,M_{\odot} < M_{\star} < 10^{12}\,M_{\odot}$. All our measurements are corrected for the effect of the inclination for disk galaxies, Galactic extinction and redshift dimming \citep[see][for details]{2020tck}. Compared to $R_{\rm e}$, the new
size measure not only captures the luminous boundary of a galaxy, but also dramatically decreases the observed scatter in the size--stellar mass relation over the five orders of magnitude in $M_{\star}$ by more than a factor of two. 
We also propose that for the very massive galaxies in our sample that have undergone merging or tidal interactions, our definition of size can potentially be used to locate the onset of their stellar haloes, i.e. a separation between the bulk of insitu and exsitu star formation in the galaxy. We explore this idea in more detail in future work. \par

\begin{figure}
    \centering
    \includegraphics[width=0.95\textwidth]{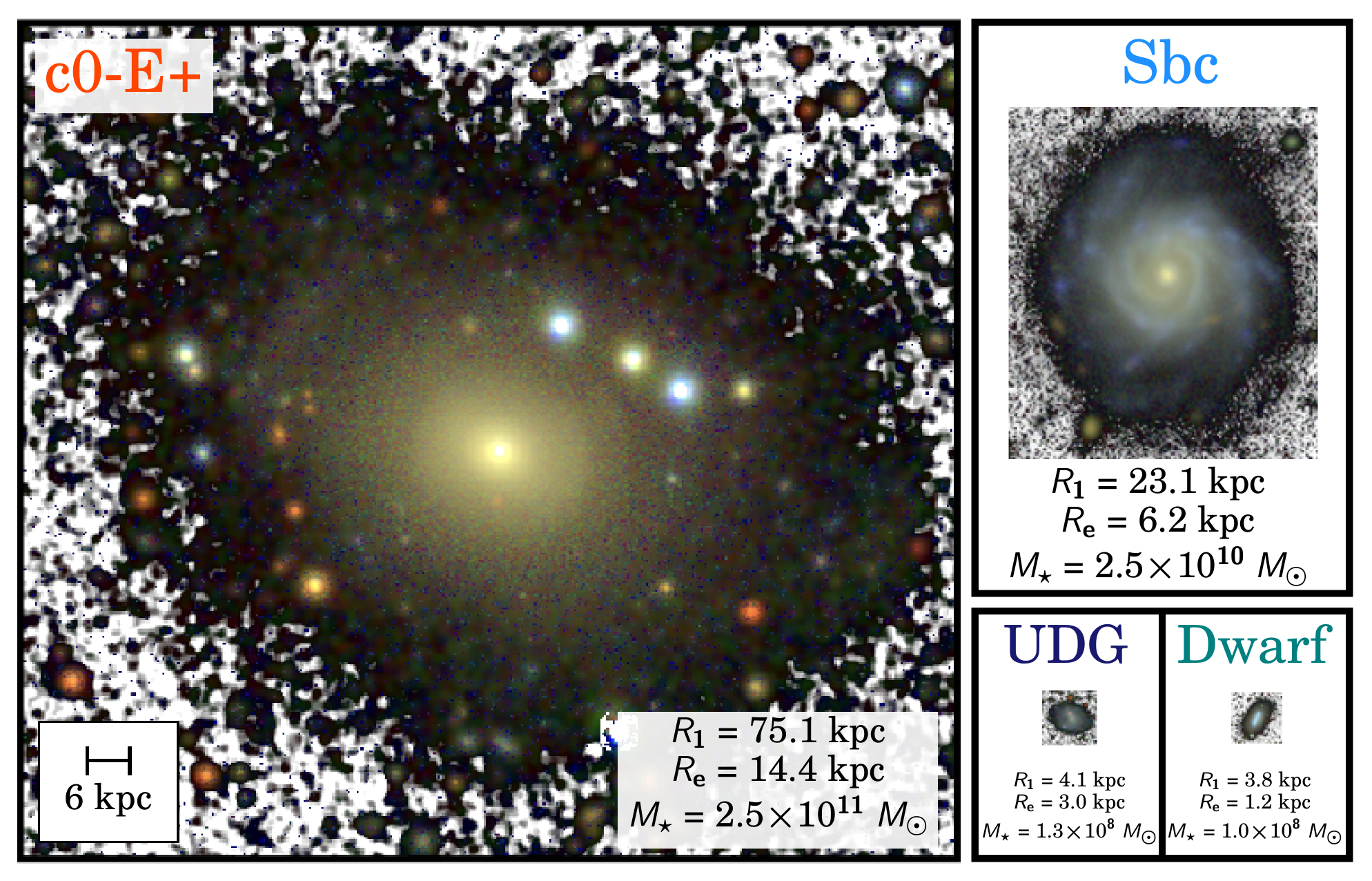}
    \caption{Representative galaxies shown to the same scale using images of the same depth ($\mu_{g, lim}$ = 29.2\,mag/arcsec$^2$ (3$\sigma;10\times10$\,arcsec$^2$)). \textit{Credit: \citet{chamba2020}}.}
    \label{fig:same_scale}
\end{figure}

The need for a more representative measure of galaxy size is further supported in \citet{chamba2020} where we demonstrated cases of a misleading use of $R_{\rm e}$ to compare the sizes of galaxies. We studied a sample of ultra-diffuse galaxies (UDGs) and dwarfs and showed that the large $R_{\rm e}$ of UDGs does not imply that the galaxy is large in size, but only that their light distributions are less concentrated. Therefore, contrary to previous reports that UDGs are Milky Way-sized, our result shows that UDGs are 10 times smaller, similar to the sizes of dwarfs (Fig. \ref{fig:same_scale}). I refer the reader to \citet{2020tck} and \citet{chamba2020} for more details. \par

\section*{Acknowledgements}

\noindent I thank Ignacio Trujillo and Johan H. Knapen for their endless support and comments which greatly improved this manuscript. I also thank Piet van der Kruit, Kenneth Freeman and John Kormendy for their insightful and critical feedback. \par 

This project has received funding from European Union's Horizon 2020 research and innovation programme under Marie Sk\l{}odowska-Curie grant agreement No. 721463 to the SUNDIAL ITN, from the State Research Agency (AEI) of the Spanish Ministry of Science and Innovation and the European Regional Development Fund (FEDER) under the grant with reference PID2019-105602GB-I00, and from IAC project P/300724, financed by the Ministry of Science and Innovation, through the State Budget and by the Canary Islands Department of Economy, Knowledge and Employment, through the Regional Budget of the Autonomous Community.

\bibliographystyle{style/apj}
\bibliography{bib.bib}

\begin{thebibliography}{}
\expandafter\ifx\csname natexlab\endcsname\relax\def\natexlab#1{#1}\fi

\bibitem[{{Arp} \& {Bertola}(1969)}]{arp1969}
{Arp}, H., \& {Bertola}, F. 1969, \apjlett, 4, 23

\bibitem[{{Azzollini} {et~al.}(2008){Azzollini}, {Trujillo}, \&
  {Beckman}}]{2008ApJ...684.1026A}
{Azzollini}, R., {Trujillo}, I., \& {Beckman}, J.~E. 2008, \apj, 684, 1026

\bibitem[{{Chamba} {et~al.}(2020){Chamba}, {Trujillo}, \&
  {Knapen}}]{chamba2020}
{Chamba}, N., {Trujillo}, I., \& {Knapen}, J.~H. 2020, A\&A, 633, L3

\bibitem[{{Crain} {et~al.}(2015){Crain}, {Schaye}, {Bower}, {Furlong},
  {Schaller}, {Theuns}, {Dalla Vecchia}, {Frenk}, {McCarthy}, {Helly},
  {Jenkins}, {Rosas-Guevara}, {White}, \& {Trayford}}]{2015crain}
{Crain}, R.~A., {Schaye}, J., {Bower}, R.~G., {et~al.} 2015, \mnras, 450, 1937

\bibitem[{{de Vaucouleurs}(1948)}]{1948AnAp...11..247D}
{de Vaucouleurs}, G. 1948, Annales d'Astrophysique, 11, 247

\bibitem[{{de Vaucouleurs}(1959{\natexlab{a}})}]{observatoireannales}
---. 1959{\natexlab{a}}, Annales de l'Observatoire du Houga, Annales de
  l'Observatoire du Houga No. v. 2, pt. 2 (Observatoire)

\bibitem[{{de Vaucouleurs}(1959{\natexlab{b}})}]{1959devucouleursb}
---. 1959{\natexlab{b}}, \aj, 64, 397

\bibitem[{{de Vaucouleurs}(1969)}]{devaucouleurs1969}
---. 1969, \aplett, 4, 17

\bibitem[{{de Vaucouleurs}(1974)}]{1974IAUSdev}
{de Vaucouleurs}, G. 1974, in IAU Symposium, Vol.~58, The Formation and
  Dynamics of Galaxies, ed. J.~R. {Shakeshaft}, 1

\bibitem[{{de Vaucouleurs}(1987)}]{1987IAUSdev}
{de Vaucouleurs}, G. 1987, in IAU Symposium, Vol. 127, Structure and Dynamics
  of Elliptical Galaxies, ed. P.~T. {de Zeeuw}, 3

\bibitem[{{de Vaucouleurs} {et~al.}(1991){de Vaucouleurs}, {de Vaucouleurs},
  {Corwin}, {Buta}, {Paturel}, \& {Fouque}}]{1991dev}
{de Vaucouleurs}, G., {de Vaucouleurs}, A., {Corwin}, Herold~G., J., {et~al.}
  1991, {Third Reference Catalogue of Bright Galaxies}

\bibitem[{{de Vaucouleurs} {et~al.}(1976){de Vaucouleurs}, {de Vaucouleurs}, \&
  {Corwin}}]{1976RC2...C......0D}
{de Vaucouleurs}, G., {de Vaucouleurs}, A., \& {Corwin}, J.~R. 1976, in Second
  reference catalogue of bright galaxies, Vol. 1976, p. Austin: University of
  Texas Press., Vol. 1976

\bibitem[{{Disney}(1976)}]{1976disney}
{Disney}, M.~J. 1976, \nat, 263, 573

\bibitem[{{Djorgovski} \& {Davis}(1987)}]{1987DS}
{Djorgovski}, S., \& {Davis}, M. 1987, \apj, 313, 59

\bibitem[{{D'Onofrio} {et~al.}(1994){D'Onofrio}, {Capaccioli}, \&
  {Caon}}]{1994donofrio}
{D'Onofrio}, M., {Capaccioli}, M., \& {Caon}, N. 1994, \mnras, 271, 523

\bibitem[{{Fish}(1963)}]{1963fish}
{Fish}, R.~A. 1963, \aj, 68, 72

\bibitem[{{Fish}(1964)}]{1964fish}
---. 1964, \apj, 139, 284

\bibitem[{{Genkin} \& {Genkina}(1970)}]{1970genkin}
{Genkin}, I.~L., \& {Genkina}, L.~M. 1970, \sovast, 14, 353

\bibitem[{{Graham}(2019)}]{2019PASA...36...35G}
{Graham}, A.~W. 2019, \pasa, 36, e035

\bibitem[{{Hall} {et~al.}(2012){Hall}, {Courteau}, {Dutton}, {McDonald}, \&
  {Zhu}}]{2012MNRAS.425.2741H}
{Hall}, M., {Courteau}, S., {Dutton}, A.~A., {McDonald}, M., \& {Zhu}, Y. 2012,
  \mnras, 425, 2741

\bibitem[{{Holmberg}(1958)}]{1958MeLuS.136....1H}
{Holmberg}, E. 1958, Meddelanden fran Lunds Astronomiska Observatorium Serie
  II, 136, 1

\bibitem[{{Hubble}(1932)}]{1932hubble}
{Hubble}, E. 1932, \apj, 76, 106

\bibitem[{{Hubble}(1926)}]{1926hubble}
{Hubble}, E.~P. 1926, \apj, 64, 321

\bibitem[{{Hubble}(1930)}]{1930hubble}
---. 1930, \apj, 71, 231

\bibitem[{{Jiang} {et~al.}(2019){Jiang}, {Dekel}, {Kneller}, {Lapiner},
  {Ceverino}, {Primack}, {Faber}, {Macci{\`o}}, {Dutton}, {Genel}, \&
  {Somerville}}]{2019MNRAS.tmp.1977J}
{Jiang}, F., {Dekel}, A., {Kneller}, O., {et~al.} 2019, \mnras, 1977

\bibitem[{{King}(1962)}]{1962king}
{King}, I. 1962, \aj, 67, 471

\bibitem[{{Kormendy}(1977)}]{1977kormendy}
{Kormendy}, J. 1977, \apj, 218, 333

\bibitem[{{Kravtsov}(2013)}]{2013ApJ...764L..31K}
{Kravtsov}, A.~V. 2013, \apjl, 764, L31

\bibitem[{{Malin} \& {Hadley}(1999)}]{1999malin}
{Malin}, D., \& {Hadley}, B. 1999, in Looking Deep in the Southern Sky, ed.
  R.~{Morganti} \& W.~J. {Couch}, 78

\bibitem[{{Mart{\'\i}nez-Lombilla} {et~al.}(2019){Mart{\'\i}nez-Lombilla},
  {Trujillo}, \& {Knapen}}]{2019MNRAS.483..664M}
{Mart{\'\i}nez-Lombilla}, C., {Trujillo}, I., \& {Knapen}, J.~H. 2019, \mnras,
  483, 664

\bibitem[{{Miller} {et~al.}(2019){Miller}, {van Dokkum}, {Mowla}, \& {van der
  Wel}}]{2019miller}
{Miller}, T.~B., {van Dokkum}, P., {Mowla}, L., \& {van der Wel}, A. 2019,
  \apjl, 872, L14

\bibitem[{{Mo} {et~al.}(1998){Mo}, {Mao}, \& {White}}]{1998MNRAS.295..319M}
{Mo}, H.~J., {Mao}, S., \& {White}, S. D.~M. 1998, \mnras, 295, 319

\bibitem[{{Mu{\~n}oz-Mateos} {et~al.}(2015){Mu{\~n}oz-Mateos}, {Sheth},
  {Regan}, {Kim}, {Laine}, {Erroz-Ferrer}, {Gil de Paz}, {Comeron}, {Hinz},
  {Laurikainen}, {Salo}, {Athanassoula}, {Bosma}, {Bouquin}, {Schinnerer},
  {Ho}, {Zaritsky}, {Gadotti}, {Madore}, {Holwerda}, {Men{\'e}ndez-Delmestre},
  {Knapen}, {Meidt}, {Querejeta}, {Mizusawa}, {Seibert}, {Laine}, \&
  {Courtois}}]{2015ApJS..219....3M}
{Mu{\~n}oz-Mateos}, J.~C., {Sheth}, K., {Regan}, M., {et~al.} 2015, \apjs, 219,
  3

\bibitem[{{Nair} {et~al.}(2011){Nair}, {van den Bergh}, \&
  {Abraham}}]{2011nair}
{Nair}, P., {van den Bergh}, S., \& {Abraham}, R.~G. 2011, \apjl, 734, L31

\bibitem[{{Pillepich} {et~al.}(2018){Pillepich}, {Springel}, {Nelson}, {Genel},
  {Naiman}, {Pakmor}, {Hernquist}, {Torrey}, {Vogelsberger}, {Weinberger}, \&
  {Marinacci}}]{2018pillepicha}
{Pillepich}, A., {Springel}, V., {Nelson}, D., {et~al.} 2018, \mnras, 473, 4077

\bibitem[{{Redman}(1936)}]{1936MNRAS..96..588R}
{Redman}, R.~O. 1936, \mnras, 96, 588

\bibitem[{{Saintonge} \& {Spekkens}(2011)}]{2011ApJ...726...77S}
{Saintonge}, A., \& {Spekkens}, K. 2011, \apj, 726, 77

\bibitem[{{S\'ersic}(1968{\natexlab{a}})}]{1968adga.book.....S}
{S\'ersic}, J.~L. 1968{\natexlab{a}}, {Atlas de Galaxias Australes}
  (Observatorio Astronomico, Universidad Nacional de Cordoba, 1968)

\bibitem[{{S\'ersic}(1968{\natexlab{b}})}]{1968sersicb}
---. 1968{\natexlab{b}}, Bulletin of the Astronomical Institutes of
  Czechoslovakia, 19, 105

\bibitem[{{Shapley}(1934)}]{1934shapley}
{Shapley}, H. 1934, Annals of Harvard College Observatory, 88, 91

\bibitem[{{Shapley}(1942)}]{1942shapley}
---. 1942, Proceedings of the National Academy of Science, 28, 186

\bibitem[{{Shen} {et~al.}(2003){Shen}, {Mo}, {White}, {Blanton}, {Kauffmann},
  {Voges}, {Brinkmann}, \& {Csabai}}]{2003shen}
{Shen}, S., {Mo}, H.~J., {White}, S. D.~M., {et~al.} 2003, \mnras, 343, 978

\bibitem[{{Sheth} {et~al.}(2010){Sheth}, {Regan}, {Hinz}, {Gil de Paz},
  {Men{\'e}ndez-Delmestre}, {Mu{\~n}oz-Mateos}, {Seibert}, {Kim},
  {Laurikainen}, {Salo}, {Gadotti}, {Laine}, {Mizusawa}, {Armus},
  {Athanassoula}, {Bosma}, {Buta}, {Capak}, {Jarrett}, {Elmegreen},
  {Elmegreen}, {Knapen}, {Koda}, {Helou}, {Ho}, {Madore}, {Masters},
  {Mobasher}, {Ogle}, {Peng}, {Schinnerer}, {Surace}, {Zaritsky},
  {Comer{\'o}n}, {de Swardt}, {Meidt}, {Kasliwal}, \&
  {Aravena}}]{2010PASP..122.1397S}
{Sheth}, K., {Regan}, M., {Hinz}, J.~L., {et~al.} 2010, \pasp, 122, 1397

\bibitem[{{Somerville} {et~al.}(2008){Somerville}, {Barden}, {Rix}, {Bell},
  {Beckwith}, {Borch}, {Caldwell}, {H{\"a}u{\ss}ler}, {Heymans}, {Jahnke},
  {Jogee}, {McIntosh}, {Meisenheimer}, {Peng}, {S{\'a}nchez}, {Wisotzki}, \&
  {Wolf}}]{2008somerville}
{Somerville}, R.~S., {Barden}, M., {Rix}, H.-W., {et~al.} 2008, \apj, 672, 776

\bibitem[{{Trujillo} {et~al.}(2020){Trujillo}, {Chamba}, \& {Knapen}}]{2020tck}
{Trujillo}, I., {Chamba}, N., \& {Knapen}, J.~H. 2020, \mnras, 493, 87

\bibitem[{{Trujillo} \& {Pohlen}(2005)}]{2005ApJ...630L..17T}
{Trujillo}, I., \& {Pohlen}, M. 2005, \apjl, 630, L17

\bibitem[{{van der Kruit}(2007)}]{2007vanderkruit}
{van der Kruit}, P.~C. 2007, \aap, 466, 883

\bibitem[{{van Dokkum} {et~al.}(2015){van Dokkum}, {Abraham}, {Merritt},
  {Zhang}, {Geha}, \& {Conroy}}]{2015giantgalaxies}
{van Dokkum}, P.~G., {Abraham}, R., {Merritt}, A., {et~al.} 2015, \apjl, 798,
  L45

\bibitem[{{Vorontsov-Vel'Yaminov}(1961)}]{1961vorontsov}
{Vorontsov-Vel'Yaminov}, B.~A. 1961, \sovast, 4, 735

\bibitem[{{York} {et~al.}(2000){York}, {Adelman}, {Anderson}, {Anderson},
  {Annis}, {Bahcall}, {Bakken}, {Barkhouser}, {Bastian}, {Berman}, {Boroski},
  {Bracker}, {Briegel}, {Briggs}, {Brinkmann}, {Brunner}, {Burles}, {Carey},
  {Carr}, {Castander}, {Chen}, {Colestock}, {Connolly}, {Crocker}, {Csabai},
  {Czarapata}, {Davis}, {Doi}, {Dombeck}, {Eisenstein}, {Ellman}, {Elms},
  {Evans}, {Fan}, {Federwitz}, {Fiscelli}, {Friedman}, {Frieman}, {Fukugita},
  {Gillespie}, {Gunn}, {Gurbani}, {de Haas}, {Haldeman}, {Harris}, {Hayes},
  {Heckman}, {Hennessy}, {Hindsley}, {Holm}, {Holmgren}, {Huang}, {Hull},
  {Husby}, {Ichikawa}, {Ichikawa}, {Ivezi{\'c}}, {Kent}, {Kim}, {Kinney},
  {Klaene}, {Kleinman}, {Kleinman}, {Knapp}, {Korienek}, {Kron}, {Kunszt},
  {Lamb}, {Lee}, {Leger}, {Limmongkol}, {Lindenmeyer}, {Long}, {Loomis},
  {Loveday}, {Lucinio}, {Lupton}, {MacKinnon}, {Mannery}, {Mantsch}, {Margon},
  {McGehee}, {McKay}, {Meiksin}, {Merelli}, {Monet}, {Munn}, {Narayanan},
  {Nash}, {Neilsen}, {Neswold}, {Newberg}, {Nichol}, {Nicinski}, {Nonino},
  {Okada}, {Okamura}, {Ostriker}, {Owen}, {Pauls}, {Peoples}, {Peterson},
  {Petravick}, {Pier}, {Pope}, {Pordes}, {Prosapio}, {Rechenmacher}, {Quinn},
  {Richards}, {Richmond}, {Rivetta}, {Rockosi}, {Ruthmansdorfer}, {Sand ford},
  {Schlegel}, {Schneider}, {Sekiguchi}, {Sergey}, {Shimasaku}, {Siegmund},
  {Smee}, {Smith}, {Snedden}, {Stone}, {Stoughton}, {Strauss}, {Stubbs},
  {SubbaRao}, {Szalay}, {Szapudi}, {Szokoly}, {Thakar}, {Tremonti}, {Tucker},
  {Uomoto}, {Vanden Berk}, {Vogeley}, {Waddell}, {Wang}, {Watanabe},
  {Weinberg}, {Yanny}, {Yasuda}, \& {SDSS Collaboration}}]{2000york}
{York}, D.~G., {Adelman}, J., {Anderson}, John~E., J., {et~al.} 2000, \aj, 120,
  1579

\end{thebibliography}

\end{document}